\documentclass[12pt,aps,prd,preprint,tightenlines,superscriptaddress,
   showpacs,nofootinbib]{revtex4-1}
\newcommand{\PRE}[1]{{#1}} 

\usepackage{bm}
\usepackage{graphicx}
\usepackage{amssymb}
\usepackage{hyperref}
\usepackage{color}
\usepackage{rotating}

\graphicspath{{figs/}}

\newcommand{\sigmaSI}{\sigma_{\text{SI}}}
\newcommand{\sigmaSD}{\sigma_{\text{SD}}}

\newcommand{\gev}{\text{GeV}}

\newcommand{\eqref}[1]{Eq.~(\ref{#1})}

\newcommand{\be}{\begin{equation}}
\newcommand{\ee}{\end{equation}}
\newcommand{\bea}{\begin{eqnarray}}
\newcommand{\eea}{\end{eqnarray}}
\newcommand{\gsim}{\lower.7ex\hbox{$\;\stackrel{\textstyle>}{\sim}\;$}}
\newcommand{\lsim}{\lower.7ex\hbox{$\;\stackrel{\textstyle<}{\sim}\;$}}

\begin{document}

\preprint{UH-511-1173-11}

\title{ \PRE{\vspace*{0.8in}}
Isospin-Violating Dark Matter in the Sun
\PRE{\vspace*{0.2in}}
}

\author{Yu Gao}
\affiliation{Department of Physics, University of Oregon, Eugene,
OR  97403, USA
\PRE{\vspace*{.1in}}
}

\author{Jason Kumar}
\affiliation{Department of Physics and Astronomy, University of
Hawai'i, Honolulu, HI 96822, USA
\PRE{\vspace*{.1in}}
}

\author{Danny Marfatia}
\affiliation{Department of Physics and Astronomy, University of Kansas,
Lawrence, KS 66045, USA
\PRE{\vspace*{.1in}}
}
\affiliation{Department of Physics, University of Wisconsin,
Madison, WI 53706, USA
\PRE{\vspace*{.1in}}
}

\begin{abstract}
\PRE{\vspace*{.3in}} We consider the prospects for studying
spin-independent isospin-violating dark matter-nucleon interactions with neutrinos from dark matter annihilation
in the Sun, with a focus on IceCube/DeepCore (IC/DC). If dark matter-nucleon interactions are isospin-violating, 
IC/DC's reach in the spin-independent cross section
may be competitive with current direct detection experiments for a
wide range of dark matter masses.
We also compare IC/DC's sensitivity to that of next generation argon, germanium, neon and xenon-based detectors.

\end{abstract}

\maketitle

\section{Introduction}
\label{sec:intro}

The IceCube Collaboration has recently completed installation
of the DeepCore extension.  An updated estimate of IceCube/DeepCore's
sensitivity to spin-dependent dark matter-nucleus scattering~\cite{bib:ic80dc}
with 180 days of data indicates that its sensitivity may be much greater than
previously expected~\cite{Braun:2009fr}.  It is therefore of interest to also consider
IC/DC's sensitivity to spin-independent scattering (see
also~\cite{Wikstrom:2009kw}).

This interest is heightened by recent developments in dark matter
model-building, which have emphasized that
dark matter couplings to protons and neutrons may be different.
In these models of isospin-violating dark matter (IVDM)~\cite{IVDM,ivdm2},
the cross section for dark matter to scatter off any isotope of an element
is determined by the relative number of protons and neutrons in that isotope.
This realization has been exploited
to construct models that can match the data from
DAMA~\cite{Bernabei:2010mq}, CoGeNT~\cite{Aalseth:2011wp} and CRESST~\cite{CRESST},
while remaining consistent with constraints
from other dark matter direct detection experiments~\cite{lightDMbounds,xe}.
In particular, if dark matter interactions
with neutrons destructively interfere with those with protons at the
$\sim 70\% $ level, then much of the low-mass data can be made consistent.

For the case of partial destructive interference, direct
detection experiments using materials with a high atomic mass number $A$
can suffer great losses
of sensitivity due to the degradation of the usual $A^2$ coherent scattering
enhancement, as well as the fact that high-$A$ materials usually have a
large neutron fraction.  Conversely, detectors utilizing low-$A$ materials, such as helium,
carbon, nitrogen, oxygen and fluorine exhibit less suppressed
sensitivity due to destructive interference.
Hydrogen has no neutrons to cause destructive interference.  A good way to study
IVDM may be through neutrino detectors~\cite{Kumar:2011hi,Chen:2011vd}, which search for the neutrino flux arising from
dark matter annihilating in the Sun after capture by elastic
scattering from solar nuclei.  Since a significant fraction of dark matter captures arise
from scattering off low-$A$ nuclei, neutrino detector sensitivity suffers the least suppression as
a result of isospin-violating interactions.

Although isospin violation has been used to understand low-mass dark
matter data, these lessons generalize to all mass ranges.
For dark matter with mass in the 30--5000 GeV range, the detection prospects
from leading experiments, such as CDMS-II and XENON100, can be significantly weakened
if dark matter interactions violate isospin.
While isospin violation also weakens the sensitivity of neutrino detectors,
these sensitivities will be much less suppressed than those of direct
detection experiments.  Among neutrino detectors, IC/DC will have the
best sensitivity to dark matter in this mass range, making it worthwhile
to consider its prospects, relative to other direct detection experiments, in probing
IVDM.

In this Letter, we perform an analysis of IC/DC's
sensitivity to the spin-independent cross section on protons $\sigmaSI^p$, including the effects of isospin violation.

\section{Indirect dark matter detection via neutrinos}

Neutrino detectors search for dark matter which is gravitationally captured
in the Sun.  The dark matter settles to the core and annihilates to Standard
Model products, which in turn produce neutrinos.  We focus on the most studied
case, where dark matter capture processes with rate $\Gamma_C$ and annihilation processes
with rate $\Gamma_A$ are in equilibrium, such that
$\Gamma_C = 2\Gamma_A$.

Neutrino detectors search for the charged leptons which are created by
incoming neutrinos through a charged-current interaction.
For the IC/DC detector we divide the muon events
into {\it upward} events (due to upward going
neutrinos interacting outside the detector volume) and {\it contained} events (due to
neutrinos that interact within the instrumented volume); see Refs.~\cite{enrico,Barger:2011em}
for details.

For both types of events,  the rate depends on dark matter interactions only
through $\Gamma_C$, and the choice of annihilation channel.
The experimental sensitivity reflects the capture rate necessary for IC/DC to distinguish the
neutrino flux due to dark matter annihilation from the atmospheric neutrino background.
The capture rate is proportional to the
dark matter-nucleon scattering cross section~\cite{DMcapture}, so we may parameterize the capture rate by
a {\it capture coefficient} $C_0$.  If dark matter-nucleon scattering is
spin-dependent, the capture coefficient is given by
\bea
\Gamma_C^{SD} (m_X) &=& \sigmaSD^p \times C_0^{SD} (m_X )\,.
\eea
On the other hand, if dark matter-nucleon scattering is spin-independent, then the effect
of coherent scattering against heavy solar nuclei depends non-trivially on the relative strength
of the dark matter couplings to neutrons and protons, $f_n$ and $f_p$, respectively.  Then,
\bea
\Gamma_C^{SI} (m_X) &=& \sigmaSI^p \times C_0^{SI} (m_X , f_n / f_p)\,.
\eea
The background event rate determines the dark matter-initiated charged lepton event rate
$\Gamma_{event}$ to which the detector is sensitive.  This in turn implies that the detector has sensitivity
to $\sigmaSI^p > \sigmaSI^{p(limit)} = \Gamma_{event}/C_0 (m_X , f_n / f_p)$.
We calculate $C_0$ using DarkSUSY~\cite{Gondolo:2004sc} assuming a local halo density
$\rho = 0.3~{\rm GeV/cm^3}$ and a Maxwellian velocity distribution with dispersion
$\bar v=270~{\rm km/s}$; see the appendix.

IC/DC reports its sensitivity to
the spin-dependent scattering cross section $\sigmaSD^p$.  From this, one can easily determine IC/DC's
sensitivity to $\sigmaSI^p$ for any choice of $f_n / f_p$ simply by rescaling the projected limit by
an appropriate ratio of capture coefficients:
\bea
\label{eq:sigmaSIsun}
\sigmaSI^{p(limit)} &=& \sigmaSD^{p(limit)} \times {C_0^{SD} (m_X) \over
C_0^{SI}(m_X ,f_n / f_p)}\,.
\eea

The IceCube Collaboration has recently presented an updated estimate for the
sensitivity of the completed IC/DC configuration to $\sigmaSD^p$ with 180 live days of data.  This estimate assumes
dark matter annihilation to the ``hard" channel, that is, to $\tau \bar \tau$ for
$m_X \leq 80~\gev$ and to $W^+ W^-$ for $m_X > 80~\gev$.  The choice of annihilation channel
affects the neutrino spectrum, which in turn affects the muon event rate, and thus the
detector's sensitivity.

A more conservative assumption would be dark matter annihilation to $b \bar b$, which is referred
to as the ``soft" channel.
The relative sensitivity in the soft channel can be obtained
by determining the ratio of muon rates at the detector (assuming a fixed dark matter-nucleon
scattering cross-section) from different annihilation channels. This procedure
is a valid approximation when the dark matter mass is much larger than the detector threshold
since the dependence on the shapes of the neutrino spectra is weakened by integrating over a wide energy range.
For dark matter masses close to the detector threshold, our results should be viewed as simply
indicative of the relative sensitivities.
The ratio of sensitivities  for
the $\tau^+\tau^-$, $W^+W^-$ and $b\bar{b}$ channels (for various $m_X$) are given in Table~\ref{tab:channels}.
\begin{table}

\begin{tabular}{c|c|c|c|c|c|c}
\hline
  & \multicolumn{2}{|c}{IceCube$_{up}$ ($>70$~GeV)} &\multicolumn{2}{|c|}{IceCube$_{con.}$ ($>70$~GeV)} &\multicolumn{2}{c}{DeepCore ($>35$~GeV)} \\
\hline
$m_X$ & \scriptsize{$W^+W^-/\tau^+\tau^-$} & $b\bar{b}/\tau^+\tau^-$&\scriptsize{$W^+W^-/\tau^+\tau^-$} &
$b\bar{b}/\tau^+\tau^-$&\scriptsize{$W^+W^-/\tau^+\tau^-$} &$b\bar{b}/\tau^+\tau^-$\\
\hline
70 &\text{---} &\text{---} &\text{---} &\text{---} &--- &4$\times 10^{-3}$ \\
82 &0 &0 &0 &0 &0.30 &5$\times 10^{-3}$ \\
90 &$7\times 10^{-6}$ &0 &$1\times 10^{-4}$ &0 &0.41 &8$\times 10^{-3}$ \\
100 &0.12 &0 &0.29 &0 &0.44 &0.012 \\
200 &0.68 &4$\times 10^{-3}$ &0.49 &0.010 &0.40 &0.044 \\
300 &0.57 &0.011 &0.39 &0.031 &0.36 &0.069 \\
400 &0.50 &0.019 &0.35 &0.052 &0.33 &0.093 \\
500 &0.45 &0.025 &0.33 &0.067 &0.32 &0.11 \\
600 &0.42 &0.031 &0.31 &0.083 &0.31 &0.12 \\
700 &0.38 &0.039 &0.30 &0.10 &0.29 &0.13 \\
800 &0.35 &0.044 &0.28 &0.11 &0.28 &0.14 \\
900 &0.35 &0.052 &0.29 &0.12 &0.29 &0.15 \\
1000 &0.32 &0.054 &0.27 &0.12 &0.28 &0.15 \\
2000 &0.27 &0.11 &0.27 &0.19 &0.27 &0.20 \\
3000 &0.26 &0.15 &0.27 &0.21 &0.28 &0.22 \\
4000 &0.22 &0.16 &0.23 &0.21 &0.23 &0.21 \\
5000 &0.41 &0.18 &0.43 &0.21 &0.43 &0.21 \\
\hline
\end{tabular}
\caption{Relative sensitivities to the $W^+W^-, b\bar{b}$ and $\tau^+\tau^-$ channels at IC/DC obtained from the
integrated muon event rates for each of these channels. The $\tau^+\tau^-$ channel proves to be most favorable provided it has
a sizable branching fraction. Due to $b-$hadron absorption by the solar medium, a DM mass significantly above the muon energy
detector threshold is necessary for the $b\bar{b}$ channel to be visible.
Both upward and contained events at IceCube assume a half-year observation time and an optimistic threshold of 70~GeV.
Experimental selection cuts are not included in the IceCube contained rate for which we assume a km$^3$ volume. The effective
area for upward events is given in Ref.~\cite{GonzalezGarcia:2009jc}. For the DeepCore effective volume we adopt the parameterization
of Ref.~\cite{Barger:2011em}.
}
\label{tab:channels}
\end{table}

\subsection{IVDM}

It is often assumed that dark matter couples identically to protons and neutrons.
Under this assumption, dark matter will scatter coherently off  nucleons in a
nucleus, leading to an $A^2$ enhancement in the
scattering cross section for heavy nuclei.  This enhancement is the reason why the solar
spin-independent capture rate is dominated by heavier nuclei, even though the Sun is
largely composed of hydrogen~\cite{DMcapture}.  While
this assumption of isospin-conserving interactions
is a valid approximation for neutralinos, it need not be true more generally.

Although isospin-violating dark matter~\cite{IVDM,ivdm2} has been used as an explanation of the
DAMA and CoGeNT data, it is really a more general scenario in which dark matter couples
differently to protons than to neutrons.  The dark matter-nucleus spin-independent
scattering cross section is given by
\bea
\sigma_A &\propto & \mu_A^2
\left[ f_p Z + f_n (A-Z) \right]^2\,,
\eea
where $\mu_A$ is the reduced mass of the dark matter-nucleus system.
The non-trivial dependence of $\sigma_A$ on $f_n / f_p$ is the reason for the
dependence of $C_0^{SI}$ on $f_n / f_p$.

\section{IceCube/DeepCore versus direct detection experiments}

The assumption of isospin-conserving
interactions {\it i.e.,} $f_n = f_p$, is commonly
made in normalizing the dark matter-nucleus scattering cross section for a nucleus with
$Z$ protons to that of dark matter scattering
against a single nucleon. This normalized cross section $\sigma_N^Z$ is given by~\cite{ivdm2}
\bea
\sigma_N^Z &=& \sigmaSI^p
\frac{\sum_i \eta_i \mu_{A_i}^2 [Z + (A_i - Z) f_n/f_p]^2}
{\sum_i \eta_i \mu_{A_i}^2 A_i^2}\,,
\eea
where $\sigmaSI^p$ is the spin-independent cross section for dark matter
to scatter off a single proton.  The summation is over the different isotopes
with atomic number $Z$, and $\eta_i$ is the natural abundance of each isotope.
As expected, $\sigma_N^Z = \sigmaSI^p$ if $f_n = f_p$, but more generally one
can have $\sigma_N^Z \ll \sigmaSI^p$.

Direct detection experiments typically
report their signals or exclusion bounds in terms of $\sigma_N^Z$.
But in the case of IVDM, it becomes necessary to compare the results of
different experiments in terms of $\sigmaSI^p$.
It is thus useful to define the ratio~\cite{ivdm2}
\bea
\label{eq:sigmaSIdd}
F_Z &\equiv& {\sigmaSI^p \over \sigma_N^Z}
=\frac{\sum_i \eta_i \mu_{A_i}^2 A_i^2}{\sum_i \eta_i \mu_{A_i}^2 [Z + (A_i - Z) f_n/f_p]^2}\,.
\eea
We may also define the quantity
\bea
F_{\odot} (m_X, f_n / f_p) &=& { C_0 (m_X , f_n / f_p =1) \over  C_0 (m_X , f_n / f_p)}\,,
\eea
which, in analogy to $F_Z$, is the factor by which a neutrino detector's sensitivity will
be suppressed if dark matter interactions violate isospin. In particular, if
$\sigmaSI^p$ is the actual dark matter-proton spin-independent scattering cross section, then
$\sigma_N^{\odot} = {\sigmaSI^p / F_{\odot}}$ is the ``normalized to nucleon" scattering
cross section which would be inferred from neutrino detector data, if one assumes isospin-conserving
interactions.

We can define the quantity $R[\odot ,Z] (m_X ,f_n /f_p)$:
\bea
R[\odot , Z] (m_X ,f_n /f_p) &\equiv& {\sigma_N^\odot \over \sigma_N^Z}
={F_Z(f_n / f_p) \over F_{\odot} (m_X, f_n / f_p) }
\nonumber\\
&\equiv& \left[R[Z, \odot ] (m_X ,f_n /f_p) \right]^{-1}\,.
\eea
For a fixed $m_X$, the maximum of $R[\odot, Z ]$ (varying over $f_n / f_p $) is the
maximum factor by which a detector with atomic number $Z$ must exclude a signal from a
neutrino detector (assuming isospin conservation) such that the signal is still excluded even
if isospin violation is allowed.
Similarly, the minimum of $R[\odot, Z ]$ (equivalently, the maximum of $R[Z, \odot  ]$) is the maximum
factor by which a neutrino detector must exclude a signal from a detector with atomic number $Z$ (assuming
isospin conservation) such that the signal is still excluded even if isospin violation is allowed.
Table~\ref{tab:rmax} shows $R_{max}[\odot ,Z ]$ for various choices of
commonly used detector elements, and various choices of $m_X$, while Table~\ref{tab:rmin} shows $R_{max}[Z, \odot]$.
Note that, for elements with only one isotope, $R_{max}[\odot ,Z] = \infty$, since the detector will be completely
insensitive to models with $f_n / f_p = -Z/(A-Z)$.  We do not list these columns in
Table~\ref{tab:rmax}.

\begin{table}
\begin{tabular}{c|ccccccc}
\hline
$m_X$  (GeV) &\text{Xe} &\text{Ge} &\text{Si} &\text{Ca} &W &\text{Ne} &C \\
\hline
10 &281 &71.2 &83.4 &80.1 &1260 &20.2 &211 \\
20 &218 &49.3 &45.2 &43.2 &1000 &10.8 &114 \\
30 &198 &42.2 &33.1 &31.5 &920 &7.82 &83.4 \\
40 &188 &39.0 &27.1 &25.6 &882 &6.37 &68.3 \\
50 &183 &37.2 &23.5 &22.1 &861 &5.51 &59.4 \\
60 &179 &35.7 &21.0 &19.7 &842 &4.92 &53.1 \\
70 &176 &34.7 &19.2 &18.0 &830 &4.50 &48.7 \\
80 &173 &34.0 &17.9 &16.8 &822 &4.19 &45.4 \\
90 &172 &33.4 &16.9 &15.8 &815 &3.95 &42.9 \\
100 &170 &33.0 &16.1 &15.0 &809 &3.76 &40.9 \\
200 &163 &30.9 &12.5 &11.6 &782 &2.92 &32.0 \\
300 &161 &30.2 &11.5 &10.5 &772 &2.66 &29.3 \\
400 &159 &29.8 &10.9 &10.0 &767 &2.54 &28.0 \\
500 &159 &29.6 &10.7 &9.76 &764 &2.47 &27.3 \\
600 &158 &29.4 &10.5 &9.59 &762 &2.43 &26.9 \\
700 &158 &29.3 &10.4 &9.47 &760 &2.40 &26.6 \\
800 &157 &29.3 &10.3 &9.39 &759 &2.37 &26.4 \\
900 &157 &29.2 &10.2 &9.33 &758 &2.36 &26.2 \\
1000 &157 &29.2 &10.2 &9.28 &757 &2.35 &26.1 \\
2000 &156 &29.0 &9.95 &9.07 &754 &2.29 &25.6 \\
3000 &156 &28.9 &9.89 &9.01 &753 &2.28 &25.5 \\
4000 &156 &28.9 &9.86 &8.98 &753 &2.27 &25.4 \\
5000 &156 &28.9 &9.84 &8.96 &753 &2.26 &25.4 \\
\hline
\end{tabular}
\caption{
$R_{max}[\odot ,Z](f_n/f_p)$ for various elements (obtained by maximizing
$R[\odot ,Z](f_n/f_p)$ over $-1\leq f_n/f_p\leq 1$).
 }
\label{tab:rmax}
\end{table}

\begin{table}
\begin{tabular}{c|ccccccccccccc}
 \hline
$m_X$ (GeV) & \text{Xe} & \text{Ge} & \text{Si} & \text{Ca} & W & \text{Ne} & C & I & \text{Cs} &
   O & \text{Na} & \text{Ar} & F \\
 \hline
 10 & 1.87 & 1.00 & 1.00 & 1.00 & 2.23 & 1.00 & 1.00 & 1.60 & 1.75 & 1.00 & 1.00 & 1.00 & 1.00 \\
 20 & 3.37 & 1.57 & 1.00 & 1.00 & 4.02 & 1.00 & 1.00 & 2.88 & 3.15 & 1.01 & 1.00 & 1.05 & 1.00 \\
 30 & 4.54 & 2.11 & 1.00 & 1.01 & 5.42 & 1.00 & 1.01 & 3.88 & 4.24 & 1.01 & 1.00 & 1.42 & 1.00 \\
 40 & 5.50 & 2.56 & 1.01 & 1.02 & 6.56 & 1.00 & 1.02 & 4.70 & 5.14 & 1.02 & 1.00 & 1.72 & 1.00 \\
 50 & 6.29 & 2.93 & 1.01 & 1.02 & 7.50 & 1.00 & 1.03 & 5.37 & 5.87 & 1.03 & 1.00 & 1.96 & 1.00 \\
 60 & 7.01 & 3.26 & 1.02 & 1.03 & 8.35 & 1.00 & 1.03 & 5.98 & 6.54 & 1.04 & 1.00 & 2.19 & 1.00 \\
 70 & 7.61 & 3.55 & 1.02 & 1.03 & 9.07 & 1.00 & 1.04 & 6.50 & 7.11 & 1.04 & 1.00 & 2.38 & 1.00 \\
 80 & 8.14 & 3.80 & 1.02 & 1.04 & 9.70 & 1.00 & 1.04 & 6.95 & 7.60 & 1.05 & 1.00 & 2.54 & 1.00 \\
 90 & 8.60 & 4.01 & 1.03 & 1.04 & 10.3 & 1.00 & 1.05 & 7.34 & 8.03 & 1.06 & 1.00 & 2.69 & 1.00 \\
 100 & 9.01 & 4.21 & 1.03 & 1.04 & 10.7 & 1.00 & 1.05 & 7.69 & 8.41 & 1.06 & 1.00 & 2.81 & 1.00 \\
 200 & 11.4 & 5.34 & 1.05 & 1.07 & 13.6 & 1.01 & 1.08 & 9.73 & 10.6 & 1.09 & 1.00 & 3.56 & 1.00 \\
 300 & 12.4 & 5.82 & 1.06 & 1.09 & 14.8 & 1.01 & 1.10 & 10.6 & 11.6 & 1.11 & 1.00 & 3.88 & 1.07 \\
 400 & 13.0 & 6.07 & 1.07 & 1.10 & 15.4 & 1.01 & 1.11 & 11.1 & 12.1 & 1.12 & 1.00 & 4.04 & 1.12 \\
 500 & 13.3 & 6.22 & 1.08 & 1.10 & 15.8 & 1.02 & 1.12 & 11.3 & 12.4 & 1.13 & 1.00 & 4.14 & 1.15 \\
 600 & 13.5 & 6.32 & 1.08 & 1.11 & 16.1 & 1.02 & 1.12 & 11.5 & 12.6 & 1.14 & 1.00 & 4.21 & 1.16 \\
 700 & 13.6 & 6.39 & 1.08 & 1.11 & 16.2 & 1.02 & 1.13 & 11.6 & 12.7 & 1.14 & 1.00 & 4.25 & 1.18 \\
 800 & 13.7 & 6.43 & 1.08 & 1.11 & 16.4 & 1.02 & 1.14 & 11.7 & 12.8 & 1.15 & 1.00 & 4.28 & 1.19 \\
 900 & 13.8 & 6.47 & 1.09 & 1.12 & 16.4 & 1.02 & 1.14 & 11.8 & 12.9 & 1.14 & 1.00 & 4.31 & 1.19 \\
 1000 & 13.9 & 6.50 & 1.09 & 1.12 & 16.5 & 1.01 & 1.14 & 11.8 & 12.9 & 1.16 & 1.00 & 4.33 & 1.20 \\
 2000 & 14.2 & 6.63 & 1.09 & 1.14 & 16.8 & 1.03 & 1.16 & 12.0 & 13.2 & 1.17 & 1.00 & 4.41 & 1.22 \\
 3000 & 14.2 & 6.66 & 1.10 & 1.14 & 16.9 & 1.03 & 1.16 & 12.1 & 13.2 & 1.18 & 1.00 & 4.43 & 1.23 \\
 4000 & 14.3 & 6.68 & 1.10 & 1.15 & 17.0 & 1.03 & 1.16 & 12.1 & 13.3 & 1.18 & 1.00 & 4.44 & 1.23 \\
 5000 & 14.3 & 6.69 & 1.10 & 1.13 & 17.0 & 1.03 & 1.17 & 12.2 & 13.3 & 1.18 & 1.00 & 4.45 & 1.23 \\
 \hline
\end{tabular}
\caption{$R_{max}[Z,\odot](f_n/f_p)$ for various elements (obtained by maximizing
$R[Z,\odot](f_n/f_p)$ over $-1\leq f_n/f_p\leq 1$).}
\label{tab:rmin}
\end{table}

As is evident from Table~\ref{tab:rmax}, isospin violation can cause direct detection
experiments to be significantly disadvantaged, relative to neutrino detectors.  This effect can be
particularly dramatic for direct detection experiments using heavy nuclei, such as xenon or tungsten.  Since heavy
atoms tend to have many more neutrons than protons, partial destructive interference between neutron and proton
interactions can strikingly reduce their sensitivity.  Partial destructive interference has less of an
effect on nuclei such as carbon, nitrogen and oxygen, which dominate the solar capture rate, and has no effect
at all on hydrogen.

On the other hand, from Table~\ref{tab:rmin} we see that neutrino detectors can never be disadvantaged by isospin
violation (relative to other direct detection experiments)
by more than a factor of $\sim 17$ within the mass range considered.  This ``worst-case scenario" occurs
when there is almost complete destructive interference ($f_n = -f_p$) in the limit of large dark matter
mass (when dark matter capture through scattering off hydrogen is very inefficient).  In this case, the detectors
which benefit the most relative to neutrino detectors are the ones with large atomic number (and thus a mismatch
between the number of protons and neutrons).

\begin{figure}[tb]
\includegraphics[width=0.95\columnwidth]{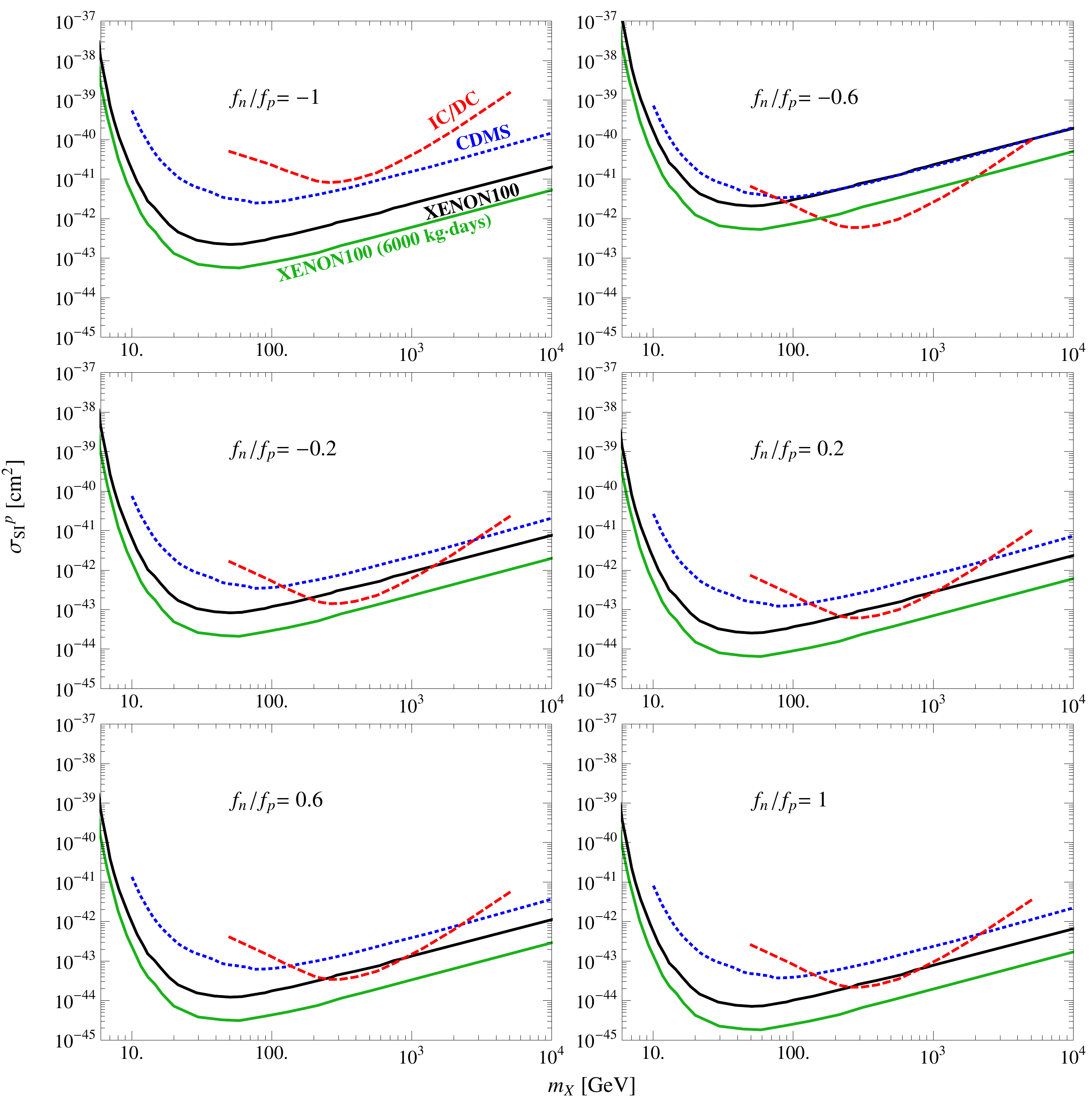}
\vspace*{-.1in}
\caption{\label{fig:sigmaSI}  Experimental sensitivity to $\sigmaSI^p$ for various choices of $f_n / f_p$, as a
function of dark matter mass $m_X$.  Current limits from CDMS-II~\cite{bib:cdms2} (blue) and
XENON100~\cite{xe} (black), expected
sensitivity for XENON100~\cite{bib:xenon100} (green),
and IceCube (80 strings) with the DeepCore extension (6
strings) in the hard channel (red), are shown.  The hard channel is annihilation to $\tau \bar \tau$ for
$m_X < 80~\gev$, and annihilation to $W^+ W^-$ for $m_X \geq 80~\gev $.}
\end{figure}

\begin{figure}[tb]
\includegraphics[scale=0.75]{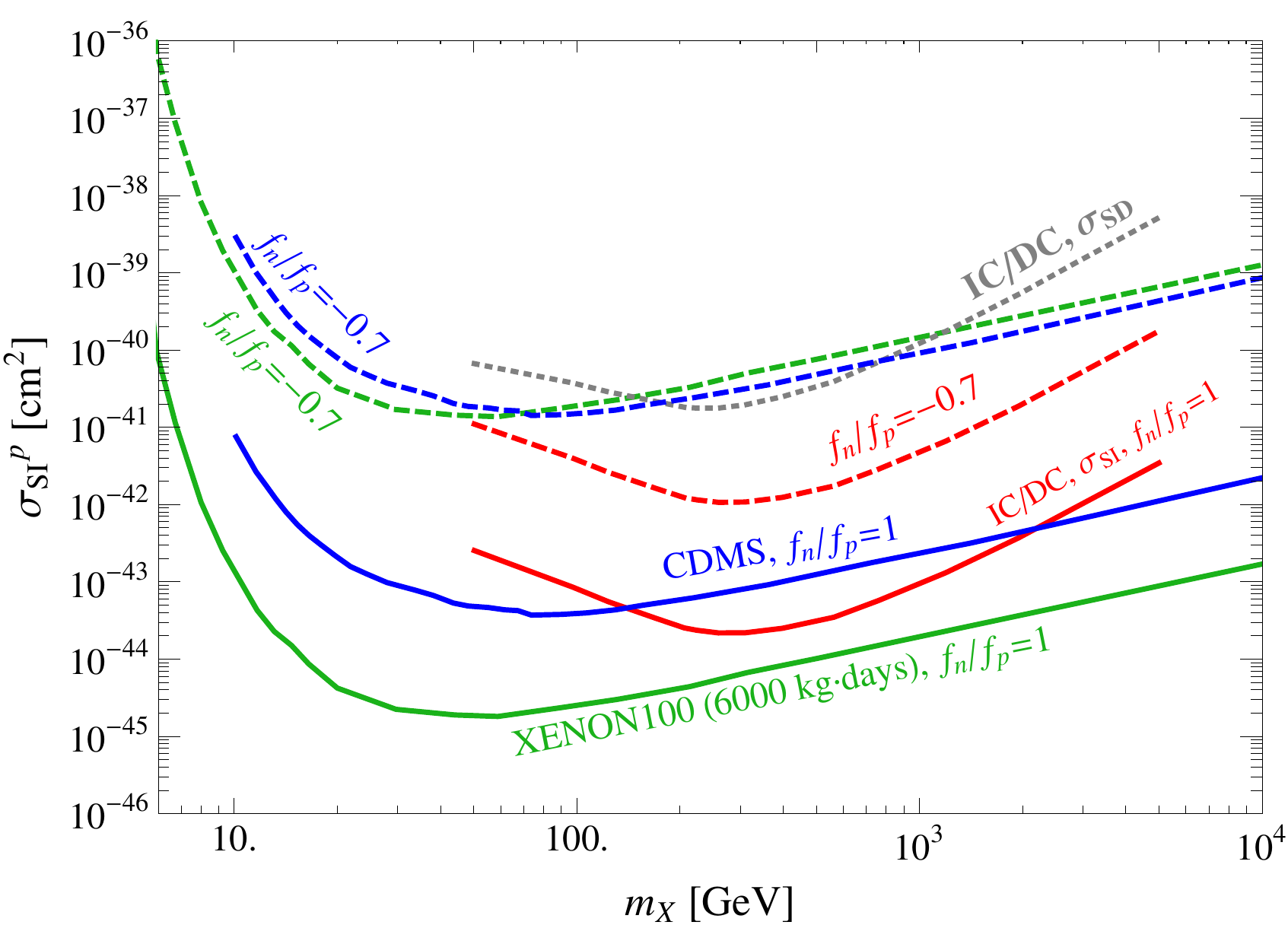}
\vspace*{-.1in}
\caption{\label{fig:sigmaSIoptimal}
Similar to Fig.~\ref{fig:sigmaSI}, for $f_n/f_p=-0.7$ (dashed curves), including current bounds from
CDMS-II (blue) and the expected sensitivity for XENON100 (green) and IC/DC (red). Solid curves
show the isospin-conserving ($f_n / f_p =1$) bounds and sensitivities for comparison. The gray dotted curve is
the expected IC/DC sensitivity to $\sigmaSD$ for the hard
channel, which is translated into a $\sigmaSI^p$ sensitivity by assuming all captures are due to SI scattering.
}
\end{figure}

\begin{figure}[tb]
\includegraphics[scale=0.38]{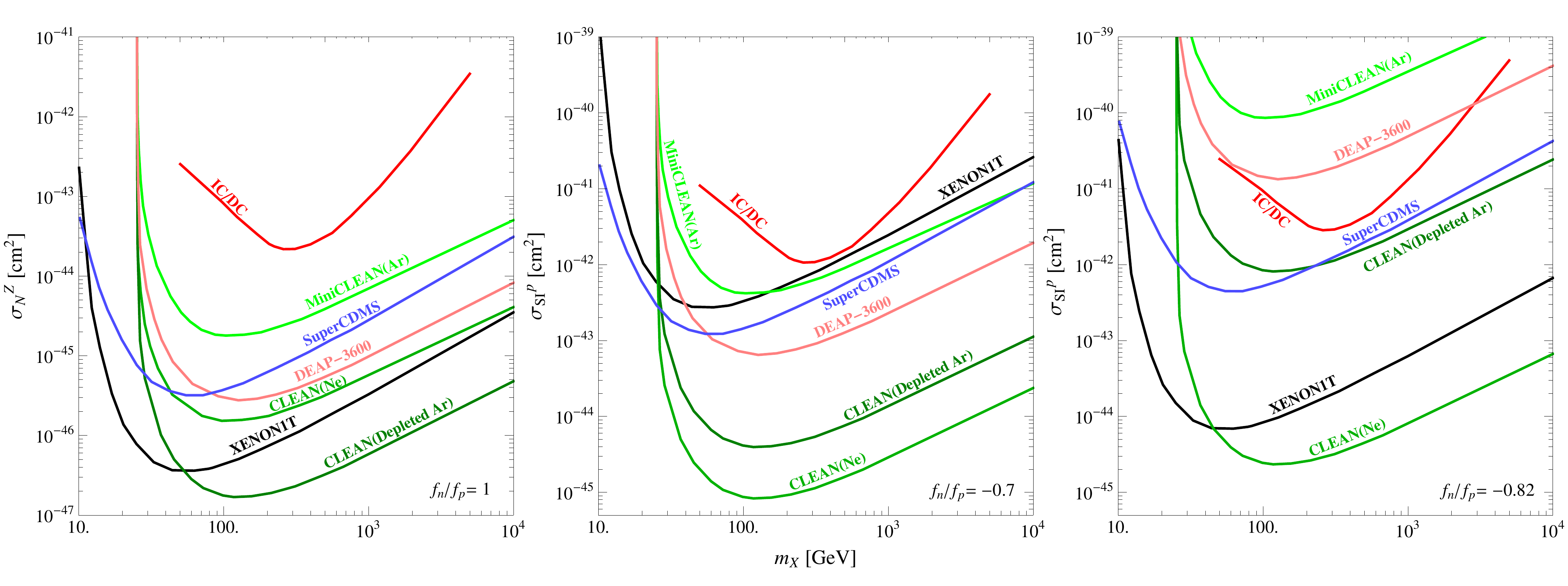}
\vspace*{-.1in}
\caption{\label{fig:sigmaSIoptimalfuture}
Experimental sensitivity to $\sigmaSI^p$ for  $f_n/f_p=1$ (left panel),
$f_n/f_p=-0.7$ (center panel) and $f_n/f_p=-0.82$ (right panel).
In addition to the expected sensitivity of IC/DC with 180 days of data we
also plot prospective bounds from XENON1T~\cite{bib:xenon100},
SuperCDMS (with a 100~kg target mass)~\cite{Brink:2005ej},
MiniCLEAN, DEAP-3600, CLEAN (Ne) and CLEAN (depleted Ar)~\cite{DEAPCLEAN} as labelled.
IC/DC's sensitivity with 1800 days
of data will be roughly three times better than that with 180 days of data.
}
\end{figure}

Using Eq.~(\ref{eq:sigmaSIsun}), one can rescale a neutrino detector's reported sensitivity
to $\sigmaSD^p$ for any annihilation channel, and determine its sensitivity to
$\sigmaSI^p$ for any choice of $f_n / f_p$ and the same annihilation channel.
Using Eq.~(\ref{eq:sigmaSIdd}), one can rescale the sensitivity to $\sigma_N^Z$ reported by
a direct detection experiment, and obtain its actual sensitivity to $\sigmaSI^p$ for any
choice of $f_n / f_p$.
In Fig.~\ref{fig:sigmaSI}, we plot  IC/DC's sensitivity (in the hard channel)
to $\sigmaSI^p$ for a variety of choices of $f_n / f_p$, assuming 180 live days of data. (In our figures,
all curves are at the 90\% C.~L.)
We also plot the CDMS-II bound, the current bound from XENON100, and the expected
sensitivity of XENON100 with $6000~{\rm kg} \cdot{\rm days}$ exposure.
In the case of isospin-conserving interactions ($f_n / f_p =1$), IC/DC's
reach is comparable with that of XENON100 in the range \mbox{$260~\gev \lesssim m_X
\lesssim 800~\gev$}.
For complete destructive interference ($f_n / f_p = -1$), current bounds are
stronger than what IC/DC can achieve over the entire mass range considered.
However, for $m_X \sim 400~\gev$ ($m_X \sim 1000~\gev$), the 180-day sensitivity of IC/DC will exceed
current XENON100 bounds
for $-0.84 \lesssim f_n / f_p \lesssim 1$ ($-0.82 \lesssim f_n / f_p \lesssim 0.28$).
Moreover, for $m_X \sim 100~\gev$, the sensitivity of IC/DC will exceed current CDMS bounds
for $-0.87 \lesssim f_n / f_p \lesssim -0.46$.
We thus see that for wide ranges of $m_X$ and $f_n / f_p$, IC/DC's sensitivity
with 180 days of data (assuming hard channel annihilation) will exceed current bounds on
dark matter-nucleon spin-independent scattering.
Note that for $f_n / f_p = -0.7$ (the value for which the sensitivity of a xenon detector is
maximally suppressed by isospin violation), the sensitivity of IC/DC exceeds current
bounds as well as the expected  sensitivity of XENON100
over the $50-5000~\gev$ mass range; see
Fig.~\ref{fig:sigmaSIoptimal}.

In Fig.~\ref{fig:sigmaSIoptimalfuture}, we show
IC/DC's estimated sensitivity with 180 days of data
compared to the projected sensitivity of several future experiments, such as XENON1T,
SuperCDMS, MiniCLEAN, DEAP-3600 and CLEAN (using both neon and depleted argon).
We plot these for $f_n / f_p = 1, -0.7, -0.82$.
$f_n / f_p =-0.82$ is the value for which an argon detector's
sensitivity is maximally suppressed.  IC/DC with 1800 days
of data will have roughly three times the sensitivity estimated with 180 days of data.

We see that for some ranges of $f_n / f_p$, IC/DC will be competitive
with XENON1T, DEAP-3600 and CLEAN (depleted Ar).  However, CLEAN (Ne) typically
will have better sensitivity than IC/DC can achieve.  This result is not
unexpected, as neon has about as many neutrons as protons, and thus a neutrino detector
will see very little relative gain in sensitivity compared to a neon detector; see Table~\ref{tab:rmax}.

\section{Conclusions}

We studied the prospects for spin-independent isospin-violating dark matter-nucleon scattering searches
at neutrino detectors, with a focus on IceCube/DeepCore using the latest estimates of
its sensitivity.

We found that isospin violation can have a very dramatic effect on the sensitivity of
neutrino detectors relative to direct detection experiments.  The ``worst-case scenario" for
neutrino detectors is complete destructive interference between proton and neutron interactions.
But even in this case, neutrino detectors cannot be disadvantaged by more than a factor $\sim 17$.
On the other hand, isospin violation can disadvantage direct detection experiments relative to
neutrino experiments by up to three orders of magnitude.  This
difference is largely due to the many different nuclei in the Sun, including the presence of hydrogen,
which is immune to the effects of destructive interference.

We plotted the expected limits from IC/DC (assuming that dark matter annihilates to the ``hard" channel)
and XENON100 and current limits
from CDMS-II and XENON100 to $\sigmaSI^p$, for a variety of choices of $f_n / f_p$.  For the standard assumption
of isospin-conserving interactions, IC/DC's projected sensitivity after 180 live days is comparable with that
of other detectors in the mass range $260~\gev \lesssim m_X \lesssim 800~\gev$.  For complete
destructive interference $f_n / f_p = -1 $, current bounds
exceed IC/DC's sensitivity.  But the most optimistic scenario for the relative sensitivity of a
neutrino detector is for $f_n / f_p \sim -0.7$; for this scenario, IC/DC's
sensitivity exceeds that of XENON100 over the entire $50-5000~\gev$ range.
It thus appears that for
this class of IVDM models, IC/DC may indeed provide the best current prospect for
dark matter detection for a wide range of parameters.

We also compared IC/DC's detection prospects with 180 days of data (its sensitivity
improves by $\sim 3$ with 1800 days of data) to that possible with upcoming direct detection
experiments, like XENON1T, SuperCDMS, and the CLEAN family of neon/argon detectors.
Although IC/DC would not be able to compete with a neon-based CLEAN detector, it
could (depending on the nature of isospin violation) provide sensitivity competitive with
the next generation of argon, germanium and xenon-based detectors.

\vskip .2in
{\bf Acknowledgments.}
We thank D.~Grant, K.~Richardson, C.~Rott, M.~Sakai and S.~Smith for useful discussions.
DM thanks the KEK Theory Center for
its hospitality during the completion of this work.
This research was supported in part by DOE
grants~DE-FG02-04ER41291, DE-FG02-04ER41308 and DE-FG02-96ER40969, and by
NSF grant PHY-0544278.

\appendix*

\section{Capture coefficients}

In Table~\ref{tab:fixSigmaP} we present $C_0^{SI} (m_X , f_n / f_p)=
\Gamma_C^{SI}(m_X, f_n / f_p) / \sigmaSI^p$ for several values of $f_n / f_p$ between $-1$ and $1$.
For values of ${f_n / f_p}$ outside this range, we instead define
$\tilde C_0^{SI} (m_X , f_p / f_n) \equiv \Gamma_C^{SI}(m_X, f_n / f_p) / \sigmaSI^n = C_0^{SI} (m_X , f_n / f_p)
\times (\sigmaSI^p / \sigmaSI^n)$.  Table~\ref{tab:fixSigmaN} presents $\tilde C_0$ in the range
$-1 \leq f_p / f_n \leq 1$.
Finally, in Table~\ref{tab:C0SD} we present $C_0^{SD} (m_X) = \Gamma_C^{SD}(m_X) / \sigmaSD^p$.

\begin{sidewaystable}
\begin{tabular}{c|cccccccccccc}
\hline
$m_X$ (GeV) &${f_n \over f_p}$=-1 &-0.8 &-0.7 &-0.6 &-0.4 &-0.2 &0 &0.2 &0.4 &0.6 &0.8 &1 \\
\hline
10 &0.10 &0.14 &0.21 &0.30 &0.58 &0.98 &1.5 &2.1 &2.9 &3.8 &4.7 &5.9 \\
20 &0.043 &0.074 &0.12 &0.20 &0.41 &0.72 &1.1 &1.6 &2.2 &2.9 &3.7 &4.5 \\
30 &0.025 &0.048 &0.087 &0.14 &0.31 &0.55 &0.86 &1.2 &1.7 &2.2 &2.8 &3.5 \\
40 &0.016 &0.035 &0.066 &0.11 &0.25 &0.44 &0.68 &0.99 &1.4 &1.8 &2.3 &2.8 \\
50 &0.012 &0.027 &0.053 &0.090 &0.20 &0.36 &0.56 &0.81 &1.1 &1.5 &1.8 &2.3 \\
60 &8.8$\times 10^{-3}$ &0.022 &0.043 &0.074 &0.17 &0.30 &0.47 &0.68 &0.93 &1.2 &1.5 &1.9 \\
70 &6.9$\times 10^{-3}$ &0.018 &0.036 &0.063 &0.14 &0.25 &0.40 &0.58 &0.79 &1.0 &1.3 &1.6 \\
80 &5.6$\times 10^{-3}$ &0.015 &0.031 &0.054 &0.12 &0.22 &0.35 &0.50 &0.69 &0.90 &1.1 &1.4 \\
90 &4.6$\times 10^{-3}$ &0.013 &0.027 &0.047 &0.11 &0.19 &0.30 &0.44 &0.60 &0.79 &1.0 &1.2 \\
100 &3.9$\times 10^{-3}$ &0.011 &0.024 &0.042 &0.095 &0.17 &0.27 &0.39 &0.54 &0.70 &0.89 &1.1 \\
200 &1.3$\times 10^{-3}$ &4.4$\times 10^{-3}$ &9.6$\times 10^{-3}$ &0.017 &0.039 &0.071 &0.11 &0.16 &0.22 &0.29 &0.37 &0.46 \\
300 &6.7$\times 10^{-4}$ &2.4$\times 10^{-3}$ &5.3$\times 10^{-3}$ &9.6$\times 10^{-3}$ &0.022 &0.040 &0.063 &0.092 &0.13 &0.17 &0.21 &0.26 \\
400 &4.2$\times 10^{-4}$ &1.5$\times 10^{-3}$ &3.4$\times 10^{-3}$ &6.1$\times 10^{-3}$ &0.014 &0.026 &0.041 &0.059 &0.081 &0.11 &0.14 &0.17 \\
500 &2.8$\times 10^{-4}$ &1.1$\times 10^{-3}$ &2.4$\times 10^{-3}$ &4.3$\times 10^{-3}$ &9.9$\times 10^{-3}$ &0.018 &0.029 &0.042 &0.057 &0.075 &0.095 &0.12 \\
600 &2.1$\times 10^{-4}$ &7.8$\times 10^{-4}$ &1.7$\times 10^{-3}$ &3.1$\times 10^{-3}$ &7.3$\times 10^{-3}$ &0.013 &0.021 &0.031 &0.042 &0.055 &0.070 &0.087 \\
700 &1.6$\times 10^{-4}$ &5.9$\times 10^{-4}$ &1.3$\times 10^{-3}$ &2.4$\times 10^{-3}$ &5.6$\times 10^{-3}$ &0.010 &0.016 &0.024 &0.032 &0.042 &0.054 &0.067 \\
800 &1.2$\times 10^{-4}$ &4.7$\times 10^{-4}$ &1.1$\times 10^{-3}$ &1.9$\times 10^{-3}$ &4.5$\times 10^{-3}$ &8.1$\times 10^{-3}$ &0.013 &0.019 &0.026 &0.034 &0.043 &0.053 \\
900 &10$\times 10^{-5}$ &3.8$\times 10^{-4}$ &8.5$\times 10^{-4}$ &1.6$\times 10^{-3}$ &3.6$\times 10^{-3}$ &6.6$\times 10^{-3}$ &0.010 &0.015 &0.021 &0.027 &0.035 &0.043 \\
1000 &8.2$\times 10^{-5}$ &3.1$\times 10^{-4}$ &7.1$\times 10^{-4}$ &1.3$\times 10^{-3}$ &3.0$\times 10^{-3}$ &5.4$\times 10^{-3}$ &8.6$\times 10^{-3}$ &0.013
&0.017 &0.023 &0.029 &0.036 \\
2000 &2.2$\times 10^{-5}$ &8.6$\times 10^{-5}$ &1.9$\times 10^{-4}$ &3.5$\times 10^{-4}$ &8.2$\times 10^{-4}$ &1.5$\times 10^{-3}$ &2.4$\times 10^{-3}$
&3.4$\times 10^{-3}$ &4.7$\times 10^{-3}$ &6.2$\times 10^{-3}$ &7.9$\times 10^{-3}$ &9.8$\times 10^{-3}$ \\
3000 &1.0$\times 10^{-5}$ &3.9$\times 10^{-5}$ &8.9$\times 10^{-5}$ &1.6$\times 10^{-4}$ &3.8$\times 10^{-4}$ &6.8$\times 10^{-4}$ &1.1$\times 10^{-3}$
&1.6$\times 10^{-3}$ &2.2$\times 10^{-3}$ &2.8$\times 10^{-3}$ &3.6$\times 10^{-3}$ &4.5$\times 10^{-3}$ \\
4000 &5.8$\times 10^{-6}$ &2.2$\times 10^{-5}$ &5.1$\times 10^{-5}$ &9.2$\times 10^{-5}$ &2.1$\times 10^{-4}$ &3.9$\times 10^{-4}$ &6.2$\times 10^{-4}$
&9.0$\times 10^{-4}$ &1.2$\times 10^{-3}$ &1.6$\times 10^{-3}$ &2.1$\times 10^{-3}$ &2.6$\times 10^{-3}$ \\
5000 &3.7$\times 10^{-6}$ &1.4$\times 10^{-5}$ &3.3$\times 10^{-5}$ &5.9$\times 10^{-5}$ &1.4$\times 10^{-4}$ &2.5$\times 10^{-4}$ &4.0$\times 10^{-4}$
&5.8$\times 10^{-4}$ &8.0$\times 10^{-4}$ &1.1$\times 10^{-3}$ &1.3$\times 10^{-3}$ &1.7$\times 10^{-3}$ \\
\hline
\end{tabular}
\normalsize
\caption{Capture coefficient $C^{SI}_0 (m_X , f_n / f_p)$  in units of $10^{29}\,\text{s}^{-1} {\rm pb}^{-1}$.\label{tab:fixSigmaP}
}
\end{sidewaystable}

\begin{sidewaystable}
\begin{tabular}{c|cccccccccccc}
\hline
\text{$m_X$ (GeV)} &${f_p \over f_n}$=-1 &-0.8 &-0.7 &-0.6 &-0.4 &-0.2 &0 &0.2 &0.4 &0.6 &0.8 &1 \\
\hline
10 &0.10 &0.14 &0.20 &0.29 &0.57 &0.96 &1.5 &2.1 &2.9 &3.7 &4.7 &5.9 \\
20 &0.043 &0.086 &0.14 &0.22 &0.44 &0.75 &1.2 &1.7 &2.2 &2.9 &3.7 &4.5 \\
30 &0.025 &0.061 &0.11 &0.17 &0.34 &0.59 &0.90 &1.3 &1.7 &2.3 &2.8 &3.5 \\
40 &0.016 &0.047 &0.083 &0.13 &0.27 &0.47 &0.72 &1.0 &1.4 &1.8 &2.3 &2.8 \\
50 &0.012 &0.038 &0.067 &0.11 &0.22 &0.38 &0.59 &0.84 &1.1 &1.5 &1.9 &2.3 \\
60 &8.8$\times 10^{-3}$ &0.031 &0.056 &0.090 &0.19 &0.32 &0.49 &0.70 &0.95 &1.2 &1.6 &1.9 \\
70 &6.9$\times 10^{-3}$ &0.026 &0.047 &0.077 &0.16 &0.27 &0.42 &0.60 &0.81 &1.1 &1.3 &1.6 \\
80 &5.6$\times 10^{-3}$ &0.022 &0.041 &0.067 &0.14 &0.24 &0.37 &0.52 &0.70 &0.91 &1.2 &1.4 \\
90 &4.6$\times 10^{-3}$ &0.019 &0.036 &0.058 &0.12 &0.21 &0.32 &0.46 &0.62 &0.80 &1.0 &1.2 \\
100 &3.9$\times 10^{-3}$ &0.017 &0.032 &0.052 &0.11 &0.19 &0.29 &0.41 &0.55 &0.71 &0.90 &1.1 \\
200 &1.3$\times 10^{-3}$ &7.0$\times 10^{-3}$ &0.013 &0.022 &0.045 &0.078 &0.12 &0.17 &0.23 &0.30 &0.38 &0.46 \\
300 &6.7$\times 10^{-4}$ &3.9$\times 10^{-3}$ &7.4$\times 10^{-3}$ &0.012 &0.026 &0.044 &0.068 &0.096 &0.13 &0.17 &0.21 &0.26 \\
400 &4.2$\times 10^{-4}$ &2.5$\times 10^{-3}$ &4.8$\times 10^{-3}$ &7.9$\times 10^{-3}$ &0.017 &0.028 &0.044 &0.062 &0.084 &0.11 &0.14 &0.17 \\
500 &2.8$\times 10^{-4}$ &1.8$\times 10^{-3}$ &3.4$\times 10^{-3}$ &5.5$\times 10^{-3}$ &0.012 &0.020 &0.031 &0.043 &0.059 &0.076 &0.096 &0.12 \\
600 &2.1$\times 10^{-4}$ &1.3$\times 10^{-3}$ &2.5$\times 10^{-3}$ &4.1$\times 10^{-3}$ &8.5$\times 10^{-3}$ &0.015 &0.023 &0.032 &0.043 &0.056 &0.071 &0.087 \\
700 &1.6$\times 10^{-4}$ &10$\times 10^{-4}$ &1.9$\times 10^{-3}$ &3.1$\times 10^{-3}$ &6.6$\times 10^{-3}$ &0.011 &0.017 &0.025 &0.033 &0.043 &0.054 &0.067 \\
800 &1.2$\times 10^{-4}$ &7.9$\times 10^{-4}$ &1.5$\times 10^{-3}$ &2.5$\times 10^{-3}$ &5.2$\times 10^{-3}$ &8.9$\times 10^{-3}$ &0.014 &0.020 &0.026 &0.034 &0.043 &0.053 \\
900 &10$\times 10^{-5}$ &6.4$\times 10^{-4}$ &1.2$\times 10^{-3}$ &2.0$\times 10^{-3}$ &4.2$\times 10^{-3}$ &7.3$\times 10^{-3}$ &0.011 &0.016 &0.021 &0.028 &0.035 &0.043 \\
1000 &8.2$\times 10^{-5}$ &5.3$\times 10^{-4}$ &1.0$\times 10^{-3}$ &1.7$\times 10^{-3}$ &3.5$\times 10^{-3}$ &6.0$\times 10^{-3}$ &9.2$\times 10^{-3}$ &0.013 &0.018 &0.023 &0.029 &0.036 \\
2000 &2.2$\times 10^{-5}$ &1.5$\times 10^{-4}$ &2.8$\times 10^{-4}$ &4.6$\times 10^{-4}$ &9.6$\times 10^{-4}$ &1.7$\times 10^{-3}$ &2.5$\times 10^{-3}$ &3.6$\times 10^{-3}$
&4.9$\times 10^{-3}$ &6.3$\times 10^{-3}$ &8.0$\times 10^{-3}$ &9.8$\times 10^{-3}$ \\
3000 &1.0$\times 10^{-5}$ &6.7$\times 10^{-5}$ &1.3$\times 10^{-4}$ &2.1$\times 10^{-4}$ &4.4$\times 10^{-4}$ &7.6$\times 10^{-4}$ &1.2$\times 10^{-3}$ &1.7$\times 10^{-3}$
&2.2$\times 10^{-3}$ &2.9$\times 10^{-3}$ &3.6$\times 10^{-3}$ &4.5$\times 10^{-3}$ \\
4000 &5.8$\times 10^{-6}$ &3.8$\times 10^{-5}$ &7.3$\times 10^{-5}$ &1.2$\times 10^{-4}$ &2.5$\times 10^{-4}$ &4.3$\times 10^{-4}$ &6.6$\times 10^{-4}$ &9.5$\times 10^{-4}$
&1.3$\times 10^{-3}$ &1.7$\times 10^{-3}$ &2.1$\times 10^{-3}$ &2.6$\times 10^{-3}$ \\
5000 &3.7$\times 10^{-6}$ &2.5$\times 10^{-5}$ &4.7$\times 10^{-5}$ &7.8$\times 10^{-5}$ &1.6$\times 10^{-4}$ &2.8$\times 10^{-4}$ &4.3$\times 10^{-4}$ &6.1$\times 10^{-4}$
&8.2$\times 10^{-4}$ &1.1$\times 10^{-3}$ &1.3$\times 10^{-3}$ &1.7$\times 10^{-3}$ \\
\hline
\end{tabular}
\normalsize
\caption{$\tilde C^{SI}_0 (m_X , f_p / f_n)$ in units of  $10^{29}\,\text{s}^{-1} {\rm pb}^{-1}$.
In comparison to Table~\ref{tab:fixSigmaP},  the contribution of hydrogen to the
capture rate is decreased, while heavy elements receive an enhancement due to the larger number of neutrons.
}
\label{tab:fixSigmaN}
\end{sidewaystable}

\begin{table}
\begin{tabular}{c|c}
\hline
\text{$m_X$} (GeV)&\text{$C^{SD}_0$} \\
\hline
10 &0.094 \\
20 &0.038 \\
30 &0.021 \\
40 &0.013 \\
50 &8.7$\times 10^{-3}$ \\
60 &6.3$\times 10^{-3}$ \\
70 &4.8$\times 10^{-3}$ \\
80 &3.8$\times 10^{-3}$ \\
90 &3.0$\times 10^{-3}$ \\
100 &2.5$\times 10^{-3}$ \\
200 &6.6$\times 10^{-4}$ \\
300 &3.0$\times 10^{-4}$ \\
400 &1.7$\times 10^{-4}$ \\
500 &1.1$\times 10^{-4}$ \\
600 &7.6$\times 10^{-5}$ \\
700 &5.6$\times 10^{-5}$ \\
800 &4.3$\times 10^{-5}$ \\
900 &3.4$\times 10^{-5}$ \\
1000 &2.7$\times 10^{-5}$ \\
2000 &6.9$\times 10^{-6}$ \\
3000 &3.1$\times 10^{-6}$ \\
4000 &1.7$\times 10^{-6}$ \\
5000 &1.1$\times 10^{-6}$ \\
\hline
\end{tabular}
\normalsize
\caption{Capture coefficient $C^{SD}_0 (m_X )$  in units of $10^{29}\,\text{s}^{-1} {\rm pb}^{-1}$.\label{tab:C0SD}}
\end{table}




\newpage
\begin{thebibliography}{99}


\bibitem{bib:ic80dc}
  C.~d.~l.~Heros [for the IceCube Collaboration],
  arXiv:1012.0184 [astro-ph.HE].

\bibitem{Braun:2009fr}
  J.~Braun and D.~Hubert [for the IceCube Collaboration],
  [arXiv:0906.1615 [astro-ph.HE]].


\bibitem{Wikstrom:2009kw}
  G.~Wikstrom and J.~Edsjo,
  JCAP {\bf 0904}, 009 (2009)
  [arXiv:0903.2986 [astro-ph.CO]].

\bibitem{IVDM}
  A.~Kurylov and M.~Kamionkowski,
  Phys.\ Rev.\  D {\bf 69}, 063503 (2004)
  [arXiv:hep-ph/0307185];
  F.~Giuliani,
  Phys.\ Rev.\ Lett.\  {\bf 95}, 101301 (2005)
  [arXiv:hep-ph/0504157];
  S.~Chang, J.~Liu, A.~Pierce, N.~Weiner and I.~Yavin,
  JCAP {\bf 1008}, 018 (2010)
  [arXiv:1004.0697 [hep-ph]].

\bibitem{ivdm2}
  J.~L.~Feng, J.~Kumar, D.~Marfatia and D.~Sanford,
  [arXiv:1102.4331 [hep-ph]].

\bibitem{Bernabei:2010mq}
  R.~Bernabei, P.~Belli, F.~Cappella, R.~Cerulli, C.~J.~Dai, A.~d'Angelo, H.~L.~He, A.~Incicchitti {\it et al.},
  Eur.\ Phys.\ J.\  {\bf C67}, 39-49 (2010)
  [arXiv:1002.1028 [astro-ph.GA]].


\bibitem{Aalseth:2011wp}
  C.~E.~Aalseth, P.~S.~Barbeau, J.~Colaresi, J.~I.~Collar, J.~D.~Leon, J.~E.~Fast, N.~Fields, T.~W.~Hossbach {\it et al.},
  [arXiv:1106.0650 [astro-ph.CO]].

\bibitem{CRESST}
See talk by W.~Seidel, {\tt
http://indico.in2p3.fr/contributionDisplay.py?contribId=195\\\&sessionId=9\&confId=1565}

\bibitem{lightDMbounds}
  D.~S.~Akerib {\it et al.} [CDMS Collaboration],
  Phys.\ Rev.\  {\bf D82}, 122004 (2010)
  [arXiv:1010.4290 [astro-ph.CO]];
  Z.~Ahmed {\it et al.} [CDMS-II Collaboration],
  Phys.\ Rev.\ Lett.\  {\bf 106}, 131302 (2011)
  [arXiv:1011.2482 [astro-ph.CO]];
  J.~Angle {\it et al.} [XENON10 Collaboration],
  [arXiv:1104.3088 [astro-ph.CO]];
  T.~Girard {\it et al.} [for the SIMPLE Collaboration],
  [arXiv:1101.1885 [astro-ph.CO]].

\bibitem{xe}
 E.~Aprile {\it et al.}  [XENON100 Collaboration],
  arXiv:1104.2549 [astro-ph.CO].

\bibitem{Kumar:2011hi}
  J.~Kumar, J.~G.~Learned, M.~Sakai and S.~Smith,
  [arXiv:1103.3270 [hep-ph]].


\bibitem{Chen:2011vd}
  S.-L.~Chen and Y.~Zhang,
  [arXiv:1106.4044 [hep-ph]].


  \bibitem{enrico}
   V.~Barger, J.~Kumar, D.~Marfatia and E.~M.~Sessolo,
  Phys.\ Rev.\  D {\bf 81}, 115010 (2010)
  [arXiv:1004.4573 [hep-ph]].

\bibitem{Barger:2011em}
  V.~Barger, Y.~Gao and D.~Marfatia,
  Phys.\ Rev.\  D {\bf 83}, 055012 (2011)
  [arXiv:1101.4410 [hep-ph]].



\bibitem{DMcapture}
  A.~Gould,
  Astrophys.\ J.\  {\bf 321}, 571 (1987);
  G.~Jungman, M.~Kamionkowski and K.~Griest,
  Phys.\ Rept.\  {\bf 267}, 195-373 (1996).
  [hep-ph/9506380].

\bibitem{Gondolo:2004sc}
  P.~Gondolo {\it et al.}
  JCAP {\bf 0407}, 008 (2004).
  [astro-ph/0406204].


\bibitem{GonzalezGarcia:2009jc}
  M.~C.~Gonzalez-Garcia, F.~Halzen and S.~Mohapatra,
  Astropart.\ Phys.\  {\bf 31}, 437 (2009)
  [arXiv:0902.1176 [astro-ph.HE]].

\bibitem{bib:cdms2}
  Z.~Ahmed {\it et al.} [The CDMS-II Collaboration],
  Science {\bf 327}, 1619-1621 (2010)
  [arXiv:0912.3592 [astro-ph.CO]];
  Z.~Ahmed {\it et al.}  [CDMS Collaboration and EDELWEISS Collaboration],
  arXiv:1105.3377 [astro-ph.CO].

\bibitem{bib:xenon100}
See talk by D. Cline, {\tt http://public.lanl.gov/friedland/info11/info11talks/ClineDM-\\INFO11.pdf}



\bibitem{Brink:2005ej}
See talk by T.~Saab,
{\tt http://indico.in2p3.fr/contributionDisplay.py?sessionId=26\&\\contribId=58\&confId=1565}

\bibitem{DEAPCLEAN}
See talk by R.~Hennings-Yeomans,
{\tt http://deapclean.org/talks/PHENO2011\_Hennings.pdf}


\end{thebibliography}
\end{document}